# Luminescence Emission from Al$_{0.3}$Ga$_{0.7}$N/GaN Multi Quantum Disc core/shell Nanowire: Numerical approach

E. Namvari[1,2], S. Shojaei[1,*], A. Asgari[1,3]

1-Photonics department, Research Institute for Applied Physics & Astronomy(RIAPA) University of Tabriz, Tabriz, Iran
2-Aras International Campus - University of Tabriz, Tabriz, Iran
3- School of Electrical, Electronic and Computer Engineering, The University of Western Australia, Crawley, WA 6009, Australia

**ABSTRACT:**

In this work, a numerical approach to investigate the room temperature luminescence emission from core/shell nanowire is presented where GaN quantum discs (QDiscs), periodically distributed in Al$_x$Ga$_{1-x}$N nanowire, is considered as core and Al$_x$Ga$_{1-x}$N as shell. Thin disk shaped (Ring shaped) n-doped region has been placed at the GaN/ Al$_x$Ga$_{1-x}$N (Al$_x$Ga$_{1-x}$N /air) interface in Al$_x$Ga$_{1-x}$N region in axial (radial) directions. To obtain energy levels and related wavefunctions, self-consistent procedure has been employed to solve Schrodinger-Poisson equations with considering the spontaneous and piezoelectric polarization. Then luminescence spectrum is studied in details to recognize the parameters influent in luminescence. The results show that the amount of doping, size of QDiscs and theirs numbers have remarkable effects on the band to band luminescence emission. Our numerical calculations gives some insights into the luminescence emission of core/shell nanowire and exhibits a useful tool to analyze findings in experiments.

* Corresponding author.
E. mail addresses: s_shojaei@tabrizu.ac.ir, shojaei.sh@gmail.com (S. Shojaei).
Tel: +98 4133392995, Fax: +98 4133347050, Skype: Saeed.shojaei

# I. Introduction

Semiconductor nanowires (NWs) represent a rapidly expanding field of research largely due to their great technological promise[1–5]. They have emerged as a platform to build complex, self-assembled and defect-free nanostructures. Several studies have recently demonstrated single nanowire (NW) Light Emitting Diodes (LEDs) and photo detectors[6–11]. During last years the epitaxial growth techniques have achieved a high degree of control of the composition, doping, and morphology of the nanowires. It has also become possible to embed axial or radial heterostructures into nanowires to engineer either carrier or phonon confinement. Doped radial core-multishell NWs have good chances to find industrial applications as high-efficiency solar cells[12–17]. On the other hand, III-nitride semiconductor nanowires containing QDiscs appear as promising building blocks for optoelectronic devices, such as photodetectors, LEDs[18,19] or optochemical sensors[20,21] and also show promise for quantum devices such as single photon emitters operating at non-cryogenic temperatures[22].

Single and multiple QDisc structures based on the $Al_xGa_{1-x}N/GaN$ material system have been extensively characterized[23–29]. However, a limited number of theoretical studies have been dedicated to the optical properties of GaN nanowires containing III-Nitride micro discs. Jacopin et.al investigated the dependence of the photoluminescence (PL) transition energy on the quantum disc thickness[30]. They concluded that 3D effective mass self-consistent Schrodinger–Poisson simulations estimates internal electric field in the QDiscs more precisely than 1D simulation. Rigutti et.al investigated Individual GaN nanowires containing $Al_xGa_{1-x}N/GaN$ QDiscs with Al content $x<16\%$ by microphotoluminescence, transmission electron microscopy, and theoretical modeling[31]. Through simulations of the quantum confinement based on a three-dimensional effective-mass model, they found the main factors influent in the spectral dispersion. However, in our knowledge no systematic theoretical study of the luminescence emission from nanowire containing nitride based multiple QDiscs with control the each parameters has been performed so far.

In the present work the luminescence emission from GaN/ $Al_xGa_{1-x}N$ multiple QDiscs embedded in single core nanowire with GaN/ $Al_xGa_{1-x}N$ shell is studied theoretically, in details. A self-consistent solution is employed for Schrodinger-Poisson equations including spontaneous and piezoelectric polarization[32]. Doping is considered within a very thin layer at the interface of discs/barriers and air/shell Band diagram and transition energies in combination with Fermi energy are calculated. Luminescence spectrum emerged from band to band transitions is calculated and analyzed. This paper is organized as follows. In Sec. II, the numerical calculations related to the self-consistent procedure in wurtzite GaN/ $Al_xGa_{1-x}N$ are described and the way to evaluate the luminescence spectrum is introduced. In Sec. III the results of numerical calculations as the energy levels and wavefunctions and luminescence emission of multiple QDiscs core/shell nanowire are presented along with detailed discussions

## II. Model and theoretical details

The model structure considered in this work is depicted in Fig.1, A series of hexagonal GaN QDiscs distributed periodically in $Al_xGa_{1-x}N$ nanowire and grown in z direction is considered as core surrounded by a lateral $Al_xGa_{1-x}N$ shell. We take the value of x as 0.3 in the following calculation. Very thin n-type doped GaN layer are inserted at the interfaces of discs/barriers, also at the interface of shell/air in radial direction. The variety of disc and barrier thicknesses in z and radial direction is considered in calculations. The nanowire ends $Al_xGa_{1-x}N$ capping layer with 5 nm thickness at each side. For simplicity, we approximate hexagonal cross section of discs and barriers with circular shape, so the cylindrical coordinate can be employed for calculations because of cylindrical symmetry of nanowire.

Surface states are neglected in this work and we are allowed to assume that a 2D electron gas are completely localized at the interface of GaN/ $Al_xGa_{1-x}N$ in GaN discs region.

The calculations have been conducted within a standard three dimensional envelope function approach in a single-band approximation. In order to employ self-consistent approach of Schrodinger-Poisson equation one needs the Kohn-Sham potential which can be written as:

$$V_{KS}(\vec{r},\vec{z}) = V(\vec{r},\vec{z}) + V_H(\vec{r},\vec{z}) + V_{XC}(\vec{r},\vec{z}) \tag{1}$$

Where V is confinement potential determined by the materials band offset in each direction that is assumed to be separable in two axial and radial directions. $V_H$ is the Hartree potential generated by free carriers and static dopants (doping is considered as n-type in our work) and is obtained by solving Poisson equation with Dirichlet boundary condition. $V_{XC}$ is approximate exchange-correlation potential that is neglected in this work[33].

Because of the cylindrical symmetry, one can write the effective-mass Schrodinger equations for electrons and holes within envelope function approximation in cylindrical coordinate separable in both z and radial axis can be written as:

$$\left[-\frac{\hbar^2}{2}\nabla_z\left[\frac{1}{m_{e(h)}^{\perp*}(\vec{z})}\nabla_z\right] + V_{ks}(\vec{z})\right]\psi_{n(m)}^{e(h)}(\vec{z}) = \epsilon_{z_{n(m)}}^{e(h)}\psi_{n(m)}^{e(h)}(\vec{z}). \tag{2}$$

$$\left[-\frac{\hbar^2}{2}\frac{1}{r}\frac{\partial}{\partial r}\left(\frac{r}{m_{e(h)}^{\parallel*}(\vec{r})}\frac{\partial}{\partial r}\right) + V_{ks}(\vec{r})\right]\psi_{n(m)}^{e(h)}(\vec{r}) = \epsilon_{r_{n(m)}}^{e(h)}\psi_{n(m)}^{e(h)}(\vec{r}). \tag{3}$$

Where $\psi_{n(m)}^{e(h)}(z), \left(\epsilon_{z_{n(m)}}^{e(h)}\right)$; and $\psi_{n(m)}^{e(h)}(r), (\epsilon_{r_{n(m)}}^{e(h)})$ are the set of Eigen functions (Eigen energies) for electrons and holes in both $\vec{z}$ and $\vec{r}$ direction, respectively. Eigen energies are calculated from linear combination of correspondent energy levels in two direction. $m_{e(h)}^{\perp}$ and $m_{e(h)}^{\parallel}$ are the coordinate dependent effective masse of electrons( holes) along the parallel and normal to z axis, respectively.

The steps of exact numerical calculation of equations of the system are explained as: (i) we set the value of Hartree potential as zero and Eigen function (Eigen energies) are obtained by considering only band offset as confining potential, (ii) finding the effect of exact

Hartree potential needs to solve the Poisson equation(Eq. 4,5) followed by calculating the density of electrons, holes (Eq.6,7) and $\rho_D$ and $\rho_A$, the completely ionized donor and acceptor densities, respectively and the effect of polarization (eq.8) obtained from equations as:

$$\nabla_{\vec{r},\vec{z}}[\varepsilon_{(\vec{r},\vec{z})}\nabla_{\vec{r},\vec{z}}V_H(\vec{r},\vec{z})] = -\frac{\rho(\vec{r},\vec{z})}{\varepsilon_0}, \qquad (4)$$

$$\rho(\vec{r},\vec{z}) = e[n_h(\vec{r},\vec{z}) - n_e(\vec{r},\vec{z}) + \rho_D(\vec{r},\vec{z}) - \rho_A(\vec{r},\vec{z})] + \sigma. \qquad (5)$$

Where

$$n_e(\vec{r},\vec{z}) = \sum_n |\psi_n^e(\vec{r},\vec{z})|^2 \frac{k_B T m_e^*}{\pi \hbar^2} \ln\left[1 + e^{\frac{\mu - \epsilon_{\vec{r},\vec{z}_n}^e}{K_B T}}\right], \qquad (6)$$

$$n_h(\vec{r},\vec{z}) = \sum_m |\psi_m^h(\vec{r},\vec{z})|^2 \frac{k_B T m_h^*}{\pi \hbar^2} \ln\left[1 + e^{\frac{\epsilon_{\vec{r},\vec{z}_m}^h - \mu}{K_B T}}\right]. \qquad (7)$$

$$\sigma = \nabla.\vec{P}. \qquad (8)$$

$T$ is the temperature and $\mu$ is Fermi level that calculated from neutral condition of charge with a numerical approach[33-36]. We impose Dirichlet boundary condition to solve Poisson equation. $\vec{P}$ is polarization at the interface of $GaN/Al_xGa_{1-x}N$[37,38]. That is caused by two sources, first the spontaneous polarization due to the polar nature of interfaces ($\vec{P}_{sp}$), and second the piezoelectric polarization due to the strain created by the lattice mismatch ($\vec{P}_{pe}$). The amount of polarization should be calculated from the blow equations:

$$\vec{P}_{tot} = \vec{P}_{sp} + \vec{P}_{pe}. \qquad (9)$$

Where from Vegard's law:

$$\vec{P}_{sp}^{AlxGa1-xN} = (1-x)\vec{P}_{sp}^{GaN} + x\vec{P}_{sp}^{AlN}. \qquad (10)$$

It is assumed that $Al_xGa_{1-x}N$ layer is relaxed, then $\vec{P}_{pe}^{AlGaN}$ is zero and all the value of piezoelectric polarization is that of GaN as[37]:

$$\vec{P}_{pe}^{GaN} = 2\frac{a-a_0}{a_0}\left(e_{31} - e_{33}\frac{C_{13}}{C_{33}}\right). \qquad (11)$$

$a_0 = a^{GaN}$[36,39] is lattice constant. $e_{31}, e_{33}, C_{13}, C_{33}$ are GaN piezoelectric constants and elastic constants of wurtzite GaN, respectively[40,41].

From Vegard's law a nonlinear interpolation for microscopic parameters such as effective mass, lattice constant and energy band gap can be written as:

$$m^*(AlxGa1-xN) = (1-x)m^*(GaN) + xm^*(AlN). \qquad (12)$$

$$a(AlxGa1-xN) = (1-x)a(GaN) + xa(AlN). \qquad (13)$$

$$E_g(Al_xGa1-xN) = xE_g(AlN) + (1-x)E_g(GaN) - bx(1-x). \tag{14}$$

x and b are the mole-fraction of Al and the bowing parameter, respectively[8,10-11]. The corresponding value of the effective parameters are listed in Table. 1.

Hartree potential obtained from step2 is used as new potential to add to band offset and solve Schrodinger equation and obtain new eigen functions(Eigen energies) then new charge density and Fermi energy. Iteration continues until the fixed value of these parameters are obtained. Details of calculation method is summarized as flowchart diagram in Fig.2. Self-consistent process continues so that fixed values of energy levels, wave functions, Fermi energy and charge density are obtained. Our calculations show very good accuracy in obtaining the values that is a privilege of our numerical procedure.

From the electron and hole states, we compute the luminescence spectra neglecting dynamic screening effects and assuming that the doping-provided carriers are in the lowest available states[33]. This means that electrons recombine from the lowest-lying conduction states with holes in states lying above the Fermi level μ corresponds to doping level.
The luminescence intensity is obtained as:

$$\tau(\omega) \propto \sum_{nm}|S_{nm}|^2 \int \frac{dk_{...}}{2\pi} \mathcal{F}^e(\epsilon^e_{nk}, T)\left(1 - \mathcal{F}^h(\epsilon^h_{mk_{...}} - \mu, T)\right)\Gamma(\epsilon^e_{nk} - \epsilon^h_{mk} - \hbar\omega - \gamma). \tag{15}$$

Where
$$S_{nm} = \int d\vec{r} \psi^e_n(\vec{r},\vec{z}) \psi^h_m(\vec{r},\vec{z}). \tag{16}$$

is the overlap integral between electron state, n, and hole state m as:
$$\psi^{e(h)}_{n(m)}(\vec{r},\vec{z}) = \psi^{e(h)}_{n(m)}(\vec{r})\psi^{e(h)}_{n(m)}(\vec{z}). \tag{17}$$

$\mathcal{F}^{e(h)}$ is the electron (hole) Fermi occupation function at a given temperature T(=300K), and Γ is a Lorentzian function with a phenomenological bandwidth γ that we set to 1 meV in order to reproduce the width at half maximum of the experimental peaks.

III. Results and discussion

In this paper, we consider the case that a nanowire contains 3 GaN QDiscs/Al$_{0.3}$Ga$_{0.7}$N barriers as core and Al$_{0.3}$Ga$_{0.7}$N as shell, is considered (as depicted in Fig.1. The width of disc and barrier are shown as $L_w$, $L_b$ (indies refer to well, barrier formed in zdirection). By solving the Schrodinger equation with taking Hartree potential as zero, three energy states obtained with values of 274, 280, 281(meV) for electrons in conduction band an three values of -3484, -3485, -3485 (meV) obtained for holes in the valence band. Panel (a) of Fig.3 shows the probability and band diagram of the structure. Panel. (b) illustrates the effect of adding the charge density(eq.5) and result of self-consistent solution of Schrodinger –Poisson equations(see Fig.2). We found the change in the value of energy levels (248, 255, 278(meV) for electrons and -3485, -3503, -3515(meV) for hole, respectively) caused by variation of band profile. From Fig.3.a, it is clear that there is tunneling probability between the coupled

QDiscs and the penetration of wave function into adjacent QDiscs is observed. Fig.3, depicts that band tilting, due to piezoelectric fields (explained in previous section) results in asymmetric wave function. Comparison between Fig.3.a and Fig.3.b shows that Hartree potential caused by doping concentration changes the probability of presence of total charge density. Furthermore, we found that the penetration of corresponding wave functions to adjacent QDiscs become weaker due to the large distance between discs (thick barrier). In fact, the barrier thickness plays very important role in configuration of charge distribution in system of periodical distributed QDiscs. In the other words, the effect of coupling of QDiscs should be considered in calculations. For example, while the probability of existence of charges (blue curve in Fig 3.a) is large in second disc, most probable presence of charge is in third disc after taking the effect of Hartree potential into account. It necessities the considering of charge densities together with barrier width in obtaining the electronic structure.

To get a detailed view, in Fig.4.a the spatial distribution of charge density (with n type donor) in the system is illustrated. It can be seen that doping positions are considered in barrier region at the interface of disc/barrier and ionization of dopants results in migration of electrons to disc region. Our findings show that electron population is enhanced in third disc illustrated in Fig.4b,c and d approves that free electron distribution (eq.5) is enhanced in third disc after taking self-consistent process. The collection of free electrons (eq.6) is pronounced in QW that is deeper. QWs depth are determined by the profile of Hartree potential (not shown here) and imposed boundary condition in addition to band offset and internal electric field of Nitride heterostructure.

We found that same trends is observed when the number of discs/barriers increases. More detailed discussions on the cases of different disc numbers will be done in next subsection.

It is worth to note that all above calculations are of z direction of cylinder and can be repeated for radial part. Hereafter we take the value of radius of disc as 4nm and that of shell as 9nm. The position of doping in radial direction is considered as 5nm from the center of the disc. The result of calculations to find energy levels and wave function for five different level of donor ionization are summarized in Fig.5 and Table2. From Table. 2, it is clear that increasing the doping concentration has not remarkable effect on energy levels, but increasing the order of doping level, modifies the energy levels considerably.

IV. Luminescence emission

In this section, luminescence emission originated by valence band to conduction band transition from a system of nanowire-multi QDisc is calculated by Eq.15. We should investigate the effect of number of quantum discs, thickness of quantum discs, distance between them (barrier thickness) and ionized donor on the luminescence emission spectrum. As mentioned in the previous section, changes in the above parameters are considered along the axes of nanowire (z axis) and parameters in radial direction is considered as constant value (radius of GaN as a core and AlGaN as a shell with $r_w = 4$nm $and r_b = 9$nm, $\rho_D = \cdots 1/cm^3$ ).

First, we compare the luminescence emission from nanowires with 3, 5, 9 and 11 QDiscsas shown in Fig.6. We take the value of barrier/well (distance between disc/disc thicknesses) and ionized doping concentration as $L_b = 2.5 nm$ $L_w = 1 nm$, $\rho_D = 1.44 \times 10^{18}$ $1/cm^3$. The height of nanowire are taken as 280, 400, 640 and 760Å, respectively. The results of numerical calculation of band profile, probability density of electron and hole presence and energy levels for both electron and hole is depicted in Fig.7 and summarized in Table. 3. As it is clear fromFig.6, with increasing the number of discs the luminescence intensity increases and its spectrum is redshifted. It is because, the transition energy between first energy levels in conduction valence band decreases as it is clear from Table. 3. Also, from Luminescence spectrum, number of picks corresponds to the relevant transitions between energy states calculations for two radial and axial direction of nanowire increases. Fig.7 presents the probability density of ground state energy levels of each quantum well that are different from each other due to the effect of Hartree potential as discussed before. A second parameter that investigated in this work is role of QDiscs thickness. To study this case, we considered 5 GaN QDiscs with different thickness (and nanowire height of 37.5, 40, 42.5, 45nm) as $L_w = 0.5, 1, 1.5$ $and$ $2 nm$, where $L_b = 2.5 nm$, $\rho_D = 1.44 \times 10^{18}$ $1/cm^3$.

The results of numerical calculation is presented in Table. 4. Band to band luminescence emission is shown in Fig.8. As it is illustrated the first band to band emission energy are increased (blueshift) and luminescence intensity decreases as the thickness of discs decreases. This trend can be explained by Fig.9 that depicts the probability of electron presence in QDiscs. As it can be seen that with increasing well width energy states shifts to lower energy and band-to-band transition energy is decreased. Also, two sets of energy states appear for holes that enhance the transition between conduction/valence band states and increases the channels of transition. From this finding, thinner QDiscs are good candidate for luminescence emission at high energies or ultraviolet region of electromagnetic spectrum . On the other hand, As the thickness is increased, the number of state increases too, thus probability of career existence is increased, then the probability of band to band recombination between electron and holes is pronounced followed by enhancement of luminescence intensity. To compare with experimental findings we consider the work of C. Rivera et.al. They investigated the Chatodoluminescence of AlGaN/GaN nanocolumnar structures. Their results show that luminesecne occurs at a higher energies for thinner quantum discs[29]. Our results are in good agreement with their finding.

The other parameter that has key role in luminescence emission is distance between adjacent QDiscs(barrier thickness, $L_b$, in axial direction band profile). To investigate the effect of this parameter, we considered a nanowire containing 5 GaN QDiscs with well width value of $L_w = 0.5 nm$, and barrier thickness of $2L_b = 1, 2, 3, 4, 5$ $and$ $6$ nm, and complete ionized donor as $\rho_D = 1.99 \times 10^{18}$ $1/cm^3$. Band to band luminescence emission is shown in Fig.11. It is clear that with increasing $L_b$, band to band emission energy increases and the amplitude of luminescence pick decreases. To discuss on the reason of this finding, the probability density profile is depicted in Fig.10 and the values of energy levels are reported in Table. 5.

From Fig.10, when QDiscs(quantum wells in axial direction) are weakly isolated from each other, they allows wave functions of electrons (holes) of adjacent disc to be overlapped.

So the probability of tunneling is increased. In contrast, thicker barrier prohibits the tunneling between discs and electrons and holes in each quantum well have much more chance to take recombination and produce luminescence. So, the nanowires containing highly packed QDiscshave efficient luminescence.

Last parameter we investigated in this work is amount of ionized donor level. Five values of doping concentration are considered. The result of numerical calculations are reported in Table.7. From Fig.12, it can be seen that increasing the doping level enhances the intensity of luminescence and redshifts that is because much more electrons are cooperated in band to band transition. We observed that high level of doping increase the Hartree potential and band bending causes the energy states of electrons to be redshifted that subsequently decrease the band to band transition energy.

V. Conclusions

In conclusion, we investigate luminescence emission properties of a core/shell nanowire composed of GaN quantum discs (QDiscs) periodically distributed in AlGaN as core and an AlGaN layer as shell. Exact numerical calculation based on Self-consistent procedure has been employed to solve Schrodinger-Poisson equations with considering the spontaneous and piezoelectric polarization. We studied how the number and thickness of discs, thickness of barrier between the discs and doping concentrations are influential in electronic structure and also luminescence emission. We found that thinner QDiscs with higher level of doping are suitable systems applicable in nanowire based emitters/detectors in ultraviolet region of electromagnetic spectrum . Our numerical approach paves the way to control the emission characteristic of core/shell nanowire composed of QDiscs.

Table 1: *Physical Parameters of Al$_x$Ga$_{(1-x)}$N at 300 K*

| Property | | unit | Ref. |
|---|---|---|---|
| **Al$_{0.3}$Ga$_{0.7}$N** | | | |
| Static dielectric constant | 10.31 | $\varepsilon_0$ | [41] |
| electron effective mass at 300K $m_e^\perp = m_e^\parallel$ | 0.20 | $m_0$ | |
| Heavy hole effective mess parallel axes z $m_h^\parallel$ | 1.87 | $m_0$ | |
| effective mess of hole in z axes in barrier $m_h^\perp$ | 3.05 | $m_0$ | |
| **wurtzite GaN** | | | |
| Band gap | 3.42 | ev | [41] |
| Static dielectric constant | 10.4 | $\varepsilon_0$ | |
| electron effective mass at 300K $m_e^\perp = m_e^\parallel$ | 0.20 | $m_0$ | [43] |
| Heavy hole effective mess parallel axese z $m_h^\parallel$ | 1.1 | $m_0$ | |
| effective mess of hole in z axese in barrier $m_h^\perp$ | 1.6 | $m_0$ | |
| $a_{GaN}$ | 3.189 | A | [39] |
| $P_{sp}^{GaN}$ | -0.029 | $c/m^2$ | |
| piezoelectric constants $e_{31}$ | -0.49 | $c/m^2$ | |
| piezoelectric constants $e_{33}$ | 0.73 | $c/m^2$ | [39] |
| elastic constants $c_{13}$ | 103 | GPa | |
| elastic constants $c_{33}$ | 405 | GPa | |
| **wurtzite AlN** | | | |
| Band gap | 6.13 | ev | [41] |
| $a_{AlN}$ | 3.112 | A | |
| $P_{sp}^{AlN}$ | -0.081 | $c/m^2$ | [39] |

*Table2: calculated energy levels for radial part for five different level of doping*

($r_w = 40\dot{A}$ and $r_b = 90\dot{A}$, $L_{dope} = 25\dot{A}$, $\rho_D = 1.44 \times 10^{18} cm^{-3}$)

| $\rho_D(1/cm^3)$ | $1.3 \times 10^{18}$ | | $1.44 \times 10^{18}$ | | $1.5 \times 10^{18}$ | | $1.65 \times 10^{18}$ | | $1.99 \times 10^{18}$ | |
|---|---|---|---|---|---|---|---|---|---|---|
| E(mev) | $\epsilon_{r_n}^e$ | $\epsilon_{r_m}^h$ | $\epsilon_{r_n}^e$ | $\epsilon_{r_m}^h$ | $\epsilon_{r_n}^e$ | $\epsilon_{r_m}^h$ | $\epsilon_{r_n}^e$ | $\epsilon_{r_m}^h$ | $\epsilon_{r_n}^e$ | $\epsilon_{r_m}^h$ |

| Description | 26.47 | -3402.01 | 26.40 | -3402.16 | 26.36 | -3402.19 | 26.29 | -3402.25 | 26.11 | 3402.40 |
|---|---|---|---|---|---|---|---|---|---|---|
| $r_w = 40 Å$ | 50.73 | -3416.51 | 104.88 | -3416.57 | 50.60 | -3416.60 | 50.51 | -3416.67 | 50.30 | -3416.81 |
| $r_b = 90 Å$ | 104.97 | -3432.37 | 50.64 | -3432.44 | 104.84 | -3432.47 | 104.75 | -3432.55 | 104.54 | -3432.72 |
| $L_{dope} = 25 Å$ | 180.59 | -3451.49 | 180.50 | -3451.58 | 180.46 | -3451.61 | 180.37 | -3451.70 | 180.15 | -3451.91 |
| | 276.24 | -3474.51 | 276.14 | -3474.61 | 276.10 | -3474.64 | 275.99 | -3474.74 | 275.75 | -3474.96 |

Table3. Calculated energy levels for four nanowire with 3, 5, 9 and 11 quantum discs.($L_b = 25 Å, L_w = 10 Å$, $\rho_D = 1.44 \times 10^{18}\ 1/cm^3$. The height of nanowire are taken as 28nm, 40nm, 60nm and 75nm, respectively)

| | | | | | | | 11** | |
|---|---|---|---|---|---|---|---|---|
| E(mev) | $\epsilon_{z_n}^e$ | $\epsilon_{z_m}^h$ | $\epsilon_{z_n}^e$ | $\epsilon_{z_m}^h$ | $\epsilon_{z_n}^e$ | $\epsilon_{z_m}^h$ | $\epsilon_{z_n}^e$ | $\epsilon_{z_m}^h$ |
| Description | 247.86 | -3485.43 | 223.84 | -3493.36 | 162.47 | -3509.99 | 120.77 | -3526.76 |
| | 255.23 | -3502.64 | 230.51 | -3513.97 | 165.47 | -3544.80 | 124.86 | -3571.84 |
| | 278.18 | -3515.50 | 231.97 | -3527.16 | 166.94 | -3562.75 | 128.44 | -3581.95 |
| | | | 243.61 | -3533.16 | 173.215 | -3563.63 | 129.73 | -3599.24 |
| $\rho_D = 1.44 \times 10^{18}\ 1/cm^3$ | | | 272.62 | -3534.92 | 183.25 | -3577.49 | 129.97 | -3600.90 |
| height | 280 | | 400 | | 640 | | 760 | |

Table4: Calculated energy levels for different thicknesses ($L_w = 5, 10, 15, 20 nm$) of five QDiscs. $L_b = 25 Å, \rho_D = 1.44 \times 10^{18}\ 1/cm^3$.

| $L_w(Å)$ | | | | | | | 20* | |
|---|---|---|---|---|---|---|---|---|
| E(mev) | $\epsilon_{z_n}^e$ | $\epsilon_{z_m}^h$ | $\epsilon_{z_n}^e$ | $\epsilon_{z_m}^h$ | $\epsilon_{z_n}^e$ | $\epsilon_{z_m}^h$ | $\epsilon_{z_n}^e$ | $\epsilon_{z_m}^h$ |
| Description | 312.55 | -3544.50 | 228.80 | -3496.97 | 152.98 | -3454.28 | 144.68 | -3406.69 |
| | 320.93 | -3563.96 | 233.18 | -3521.18 | 158.08 | -3479.81 | 162.83 | -3419.06 |
| | 326.39 | -3567.03 | 234.98 | -3534.19 | 166.25 | -3494.72 | 171.76 | -3427.09 |
| $\rho_D = 1.44 \times 10^{18}\ 1/cm^3$ | 334.32 | -3573.12 | 247.40 | -3536.11 | 180.97 | -3502.99 | 179.75 | -3435.82 |
| | 347.41 | -3580.65 | 271.36 | -3538.73 | 206.18 | -3508.04 | 191.8 | -3453.07 |
| height | 357 | | 400 | | 425 | | 450 | |

Table5: Calculated energy levels for QDiscs with different barrier thicknesses of $2L_b = 10, 20, 30, 40, 50\ and\ 60\ Å$, $L_w = 5 Å$, and ionized donor level as $\rho_D = 1.99 \times 10^{18}\ 1/cm^3$

| $L_b(Å)$ | 5 | | 10 | | 15 | | 20 | | 25 | | 30 | |
|---|---|---|---|---|---|---|---|---|---|---|---|---|
| E(mev) | $\epsilon_{z_n}^e$ | $\epsilon_{z_m}^h$ | $\epsilon_{z_n}^e$ | $\epsilon_{z_m}^h$ | $\epsilon_{z_n}^e$ | $\epsilon_{z_m}^h$ | $\epsilon_{z_n}^e$ | $\epsilon_{z_m}^h$ | $\epsilon_{z_n}^e$ | $\epsilon_{z_m}^h$ | $\epsilon_{z_n}^e$ | $\epsilon_{z_m}^h$ |
| Description $L_w = 40 Å$ $L_{dope} = 5 Å$ $\rho_D = 1.99 \times 10^{18}\ 1/cm^3$ | 122.11 | -3366.15 | 120.21 | -3395.24 | 122.95 | -3408.46 | 122.61 | -3416.23 | 125.04 | -3416.97 | 124.43 | -3429.21 |
| | 159.80 | -3391.67 | 143.51 | -3420.08 | 137.11 | -3429.00 | 138.93 | -3437.71 | 143.81 | -3438.54 | 134.31 | -3449.31 |
| | 187.06 | -3412.35 | 160.69 | -3440.25 | 150.82 | -3444.28 | 148.25 | -3449.39 | 145.40 | -3444.43 | 146.36 | -3461.65 |
| | 219.21 | -3430.54 | 176.91 | -3456.23 | 166.04 | -3457.70 | 159.74 | -3458.59 | 160.72 | -3455.51 | 147.19 | -3462.44 |
| | 256.61 | -3462.14 | 207.08 | -3477.60 | 193.58 | -3478.32 | 180.74 | -3474.15 | 172.16 | -3464.01 | 166.84 | -3471.86 |
| height | 250 | | 300 | | 350 | | 400 | | 450 | | 500 | |

Table6: calculated energy levels for QDiscs with different doping level.

| $\rho_D (1/cm^3)$ | | | | | | | | | | |
|---|---|---|---|---|---|---|---|---|---|---|
| E(mev) | $\epsilon_{z_n}^e$ | $\epsilon_{z_m}^h$ | $\epsilon_{z_n}^e$ | $\epsilon_{z_m}^h$ | $\epsilon_{z_n}^e$ | $\epsilon_{z_m}^h$ | $\epsilon_{z_n}^e$ | $\epsilon_{z_m}^h$ | $\epsilon_{z_n}^e$ | $\epsilon_{z_m}^h$ |

| Description | 151.13 | -3390.60 | 147.70 | 3393.71 | 145.61 | -3395.56 | 140.62 | -3400.12 | 129.59 | -3410.20 |
|---|---|---|---|---|---|---|---|---|---|---|
| | 173.31 | -3398.41 | 168.87 | -3402.80 | 166.17 | -3405.43 | 159.70 | -3411.87 | 145.35 | -3426.11 |
| | 182.90 | -3404.82 | 178.16 | -3409.56 | 175.30 | -3412.41 | 168.40 | -3419.32 | 153.11 | -3434.61 |
| | 189.35 | -3414.09 | 184.93 | -3418.57 | 182.29 | -3421.28 | 175.81 | -3427.78 | 161.48 | -3442.21 |
| Height=450Å | 197.05 | -3434.96 | 193.87 | -3438.49 | 191.97 | -3440.64 | 187.30 | -3445.77 | 176.95 | -3457.16 |

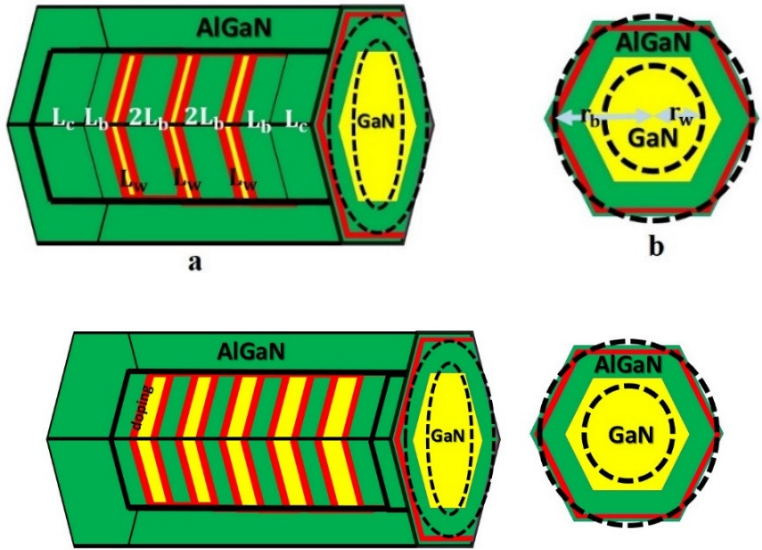

*Fig.1. Hybrid Semiconductor Nanowire –Qdiscs with AlGaN( green )and GaN (yellow). Doped region is illustrated as red color.*

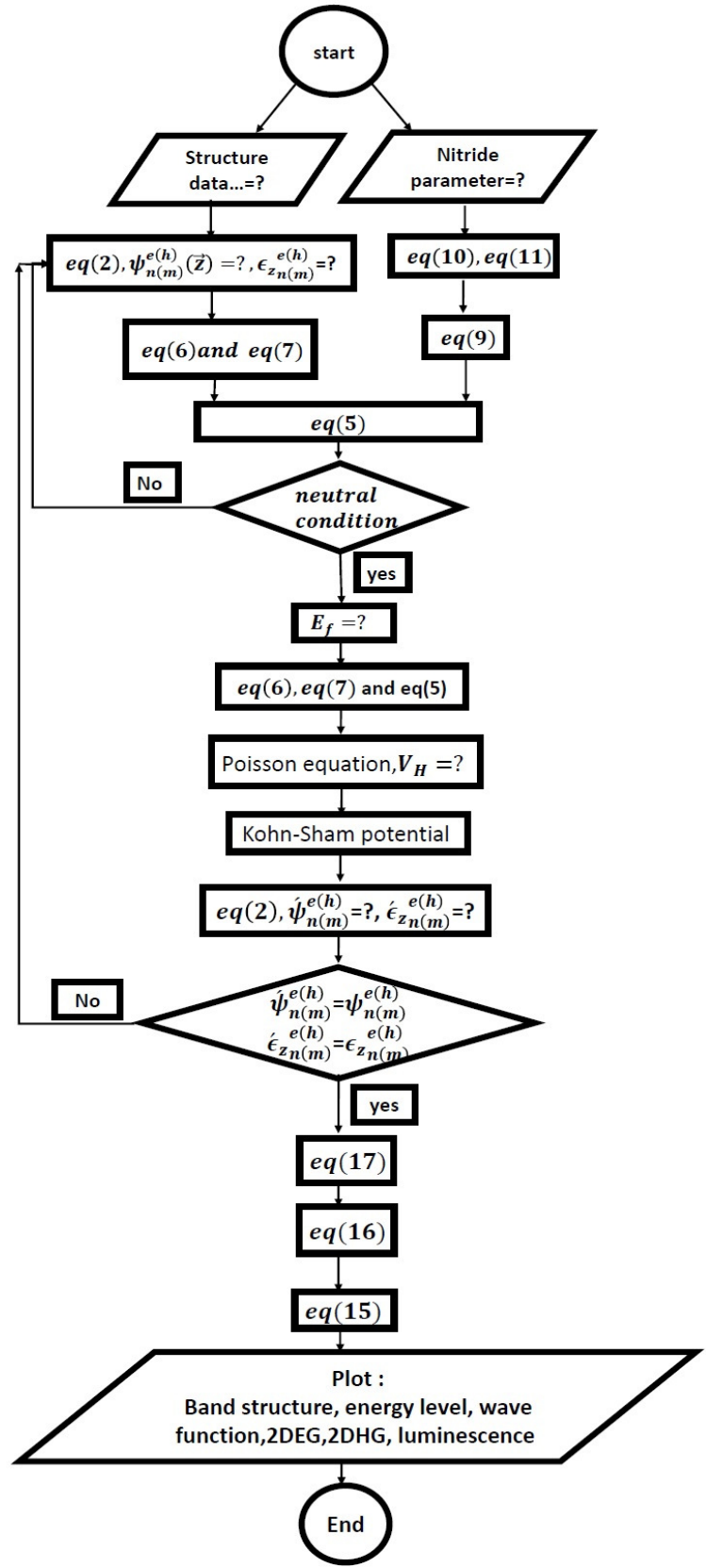

*Fig.2. Flowchart diagram of self-consistent calculations.*

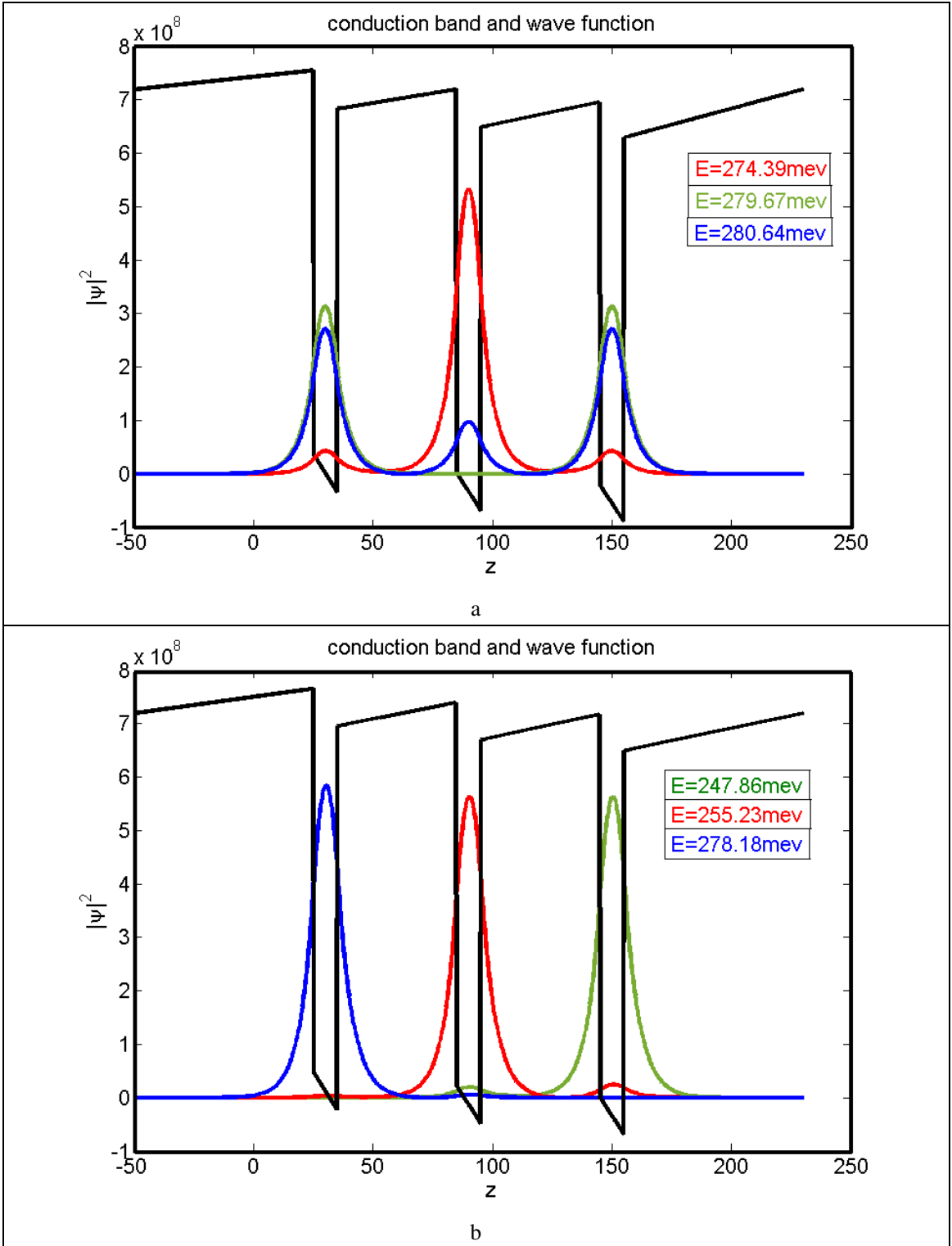

*Fig.3. Probability density in conduction band a) before , b) after the self-consisten calculation. 3 GaN Qdiscs with $L_w = 10Å$ is considered among AlGaN with the thickness of $L_b = 25Å$ and $L_c = 50Å$ is thickness of capping layer*

$$\rho_D = 1.44 \times 10^{18} cm^{-3}.$$

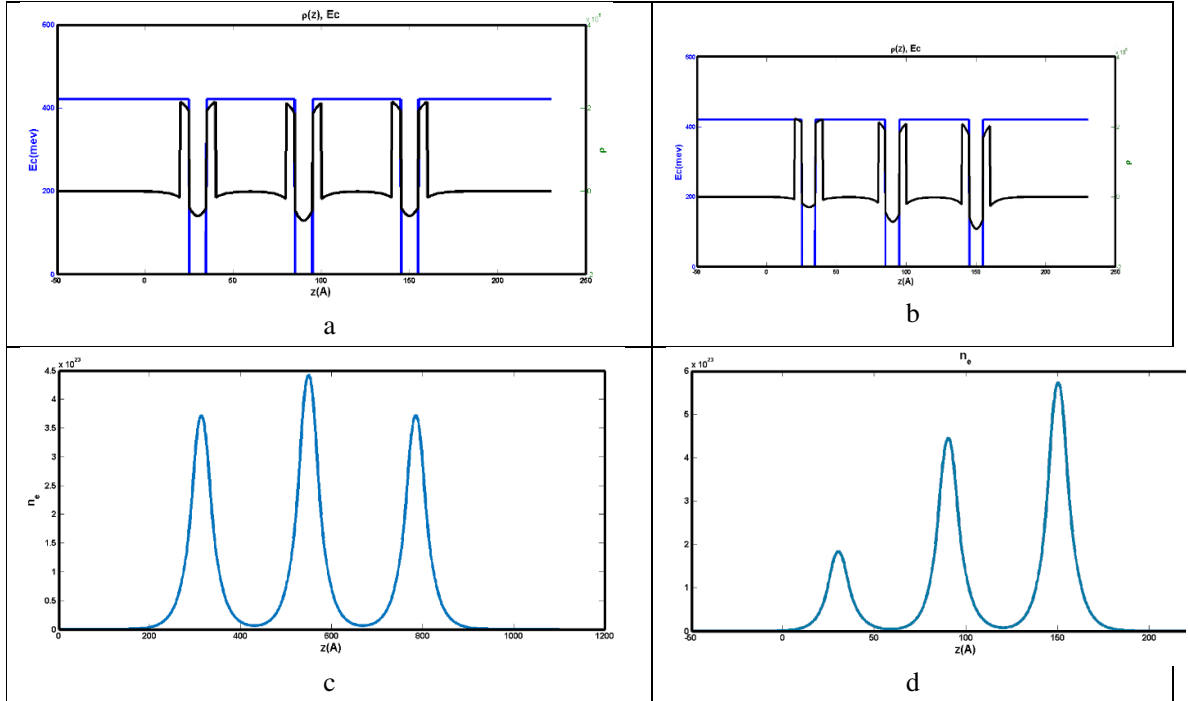

*Fig.4. The volumetric total charge density ($1/m^3$), left and right column are the results of calculations before and after self-consistent calculations, respectively. ($L_w = 10\dot{A}$, $L_b = 25\dot{A}$, $2L_b = 50\dot{A}$, $L_c = 50\dot{A}$, $L_{dope} = 5\dot{A}$, $\rho_D = 1.44 \times 10^{18} cm^{-3}$)*

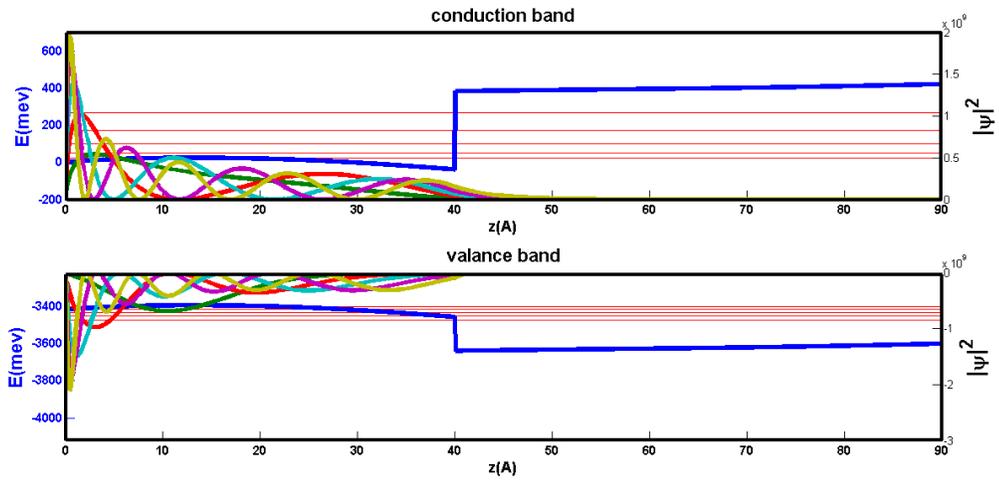

*Fig.5: Wave functions and band diagram along the radial part of nanowire ($r_w = 40\dot{A}$ and $r_b = 90\dot{A}$, $L_{dope} = 25\dot{A}$, $\rho_D = 1.44 \times 10^{18} cm^{-3}$)*

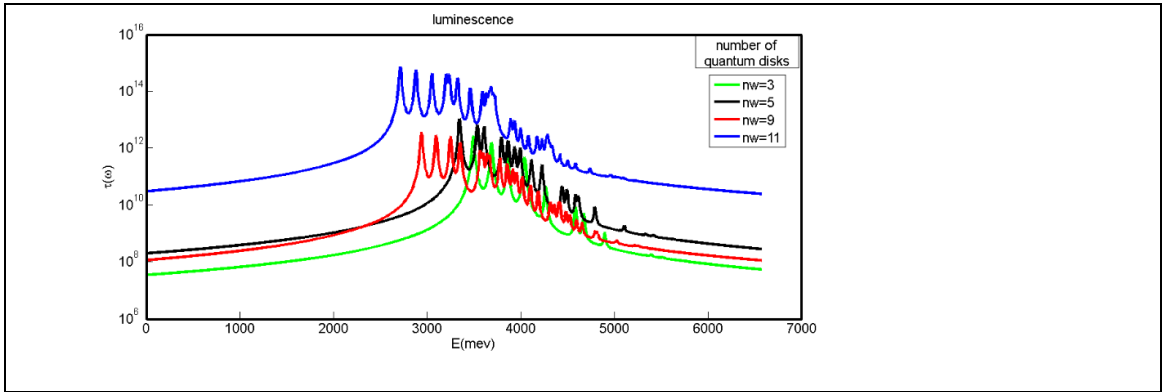

*Fig.6. Luminescence spectrum for 3, 5, 9 and 11 GaN QDiscs embedded in AlGaN. ($L_b = 25Å, L_w = 10Å$, $\rho_D = 1.44 \times 10^{18}\ 1/cm^3$. The height of nanowire are taken as 280, 400, 640 and 760Å, respectively).*

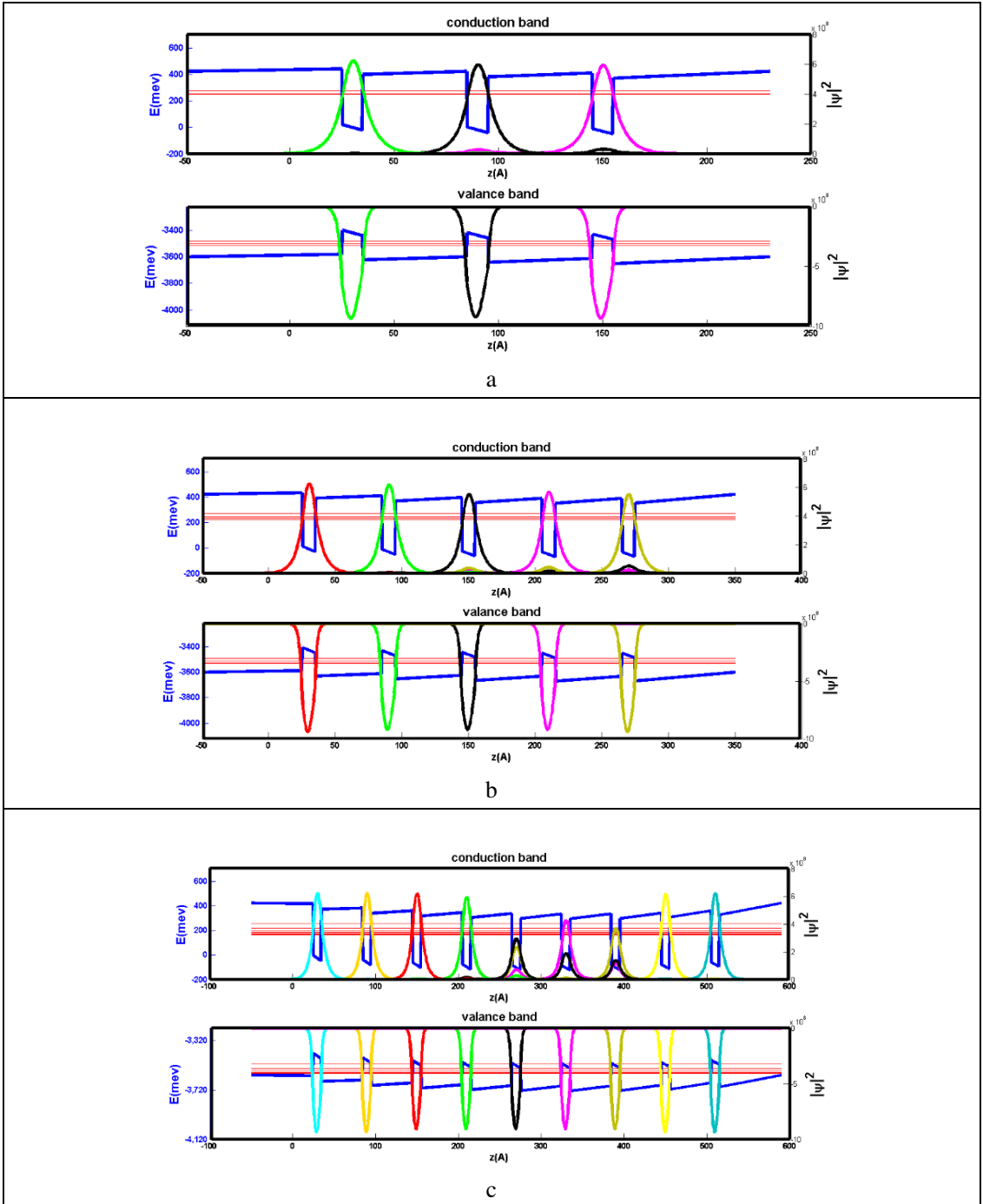

Fig.7: Band diagram and wave function for nanowires with a)3, b)5, c)9 and d)11 QDiscs. ($L_b = 25 \dot{A}, L_w = 10 \dot{A}$, $\rho_D = 1.44 \times 10^{18}$ $1/cm^3$.

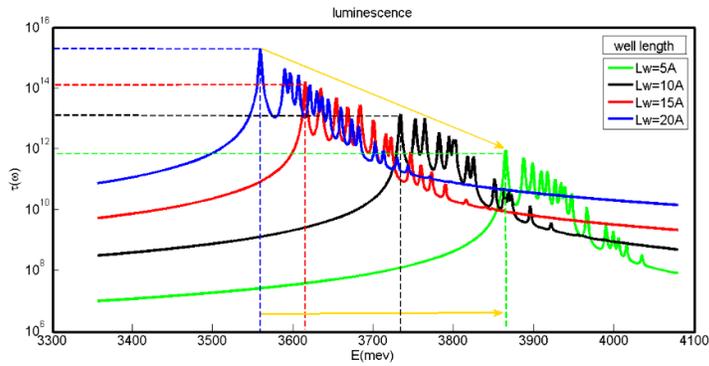

*Fig.8. Luminescence emission for 5 GaN QDiscs with different thickness 5, 10. 15 and 20 Å embedded in AlGaN. (The height of 375, 400, 425, 450Å, $L_b = 25Å, \rho_D = 1.44 \times 10^{18}\ 1/cm^3$).*

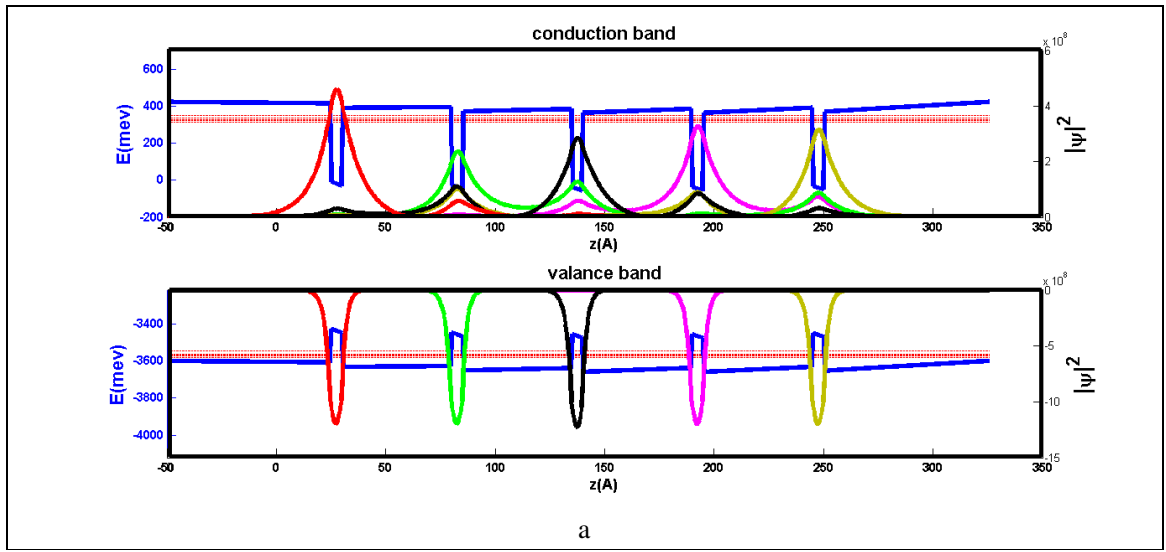

a

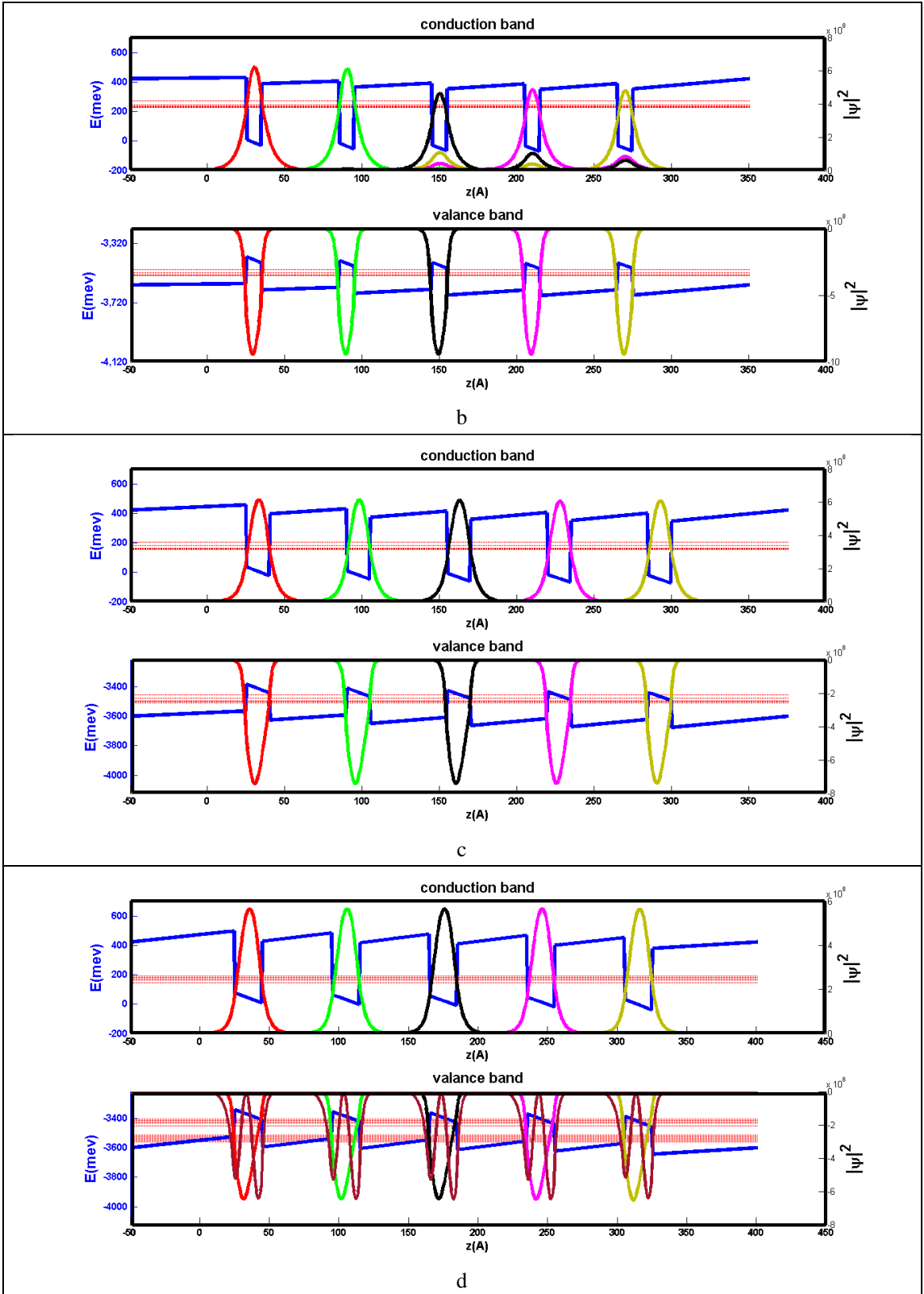

*Fig.9: wave function and band diagram for nanowire with different thickness a) $L_w$ =5, b) $L_w$ =10, c) $L_w$ =15 and d) $L_w$ =20 Å. (The height of 375, 400, 425, 450Å, $L_b = 25Å, \rho_D = 1.44 \times 10^{18}\ 1/cm^3$)*

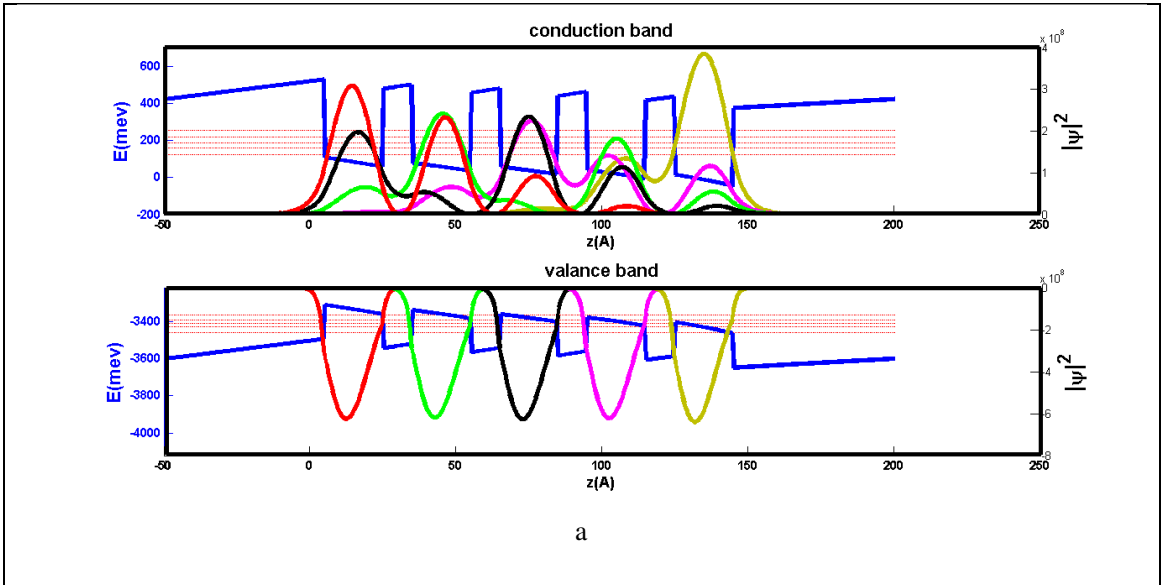

a

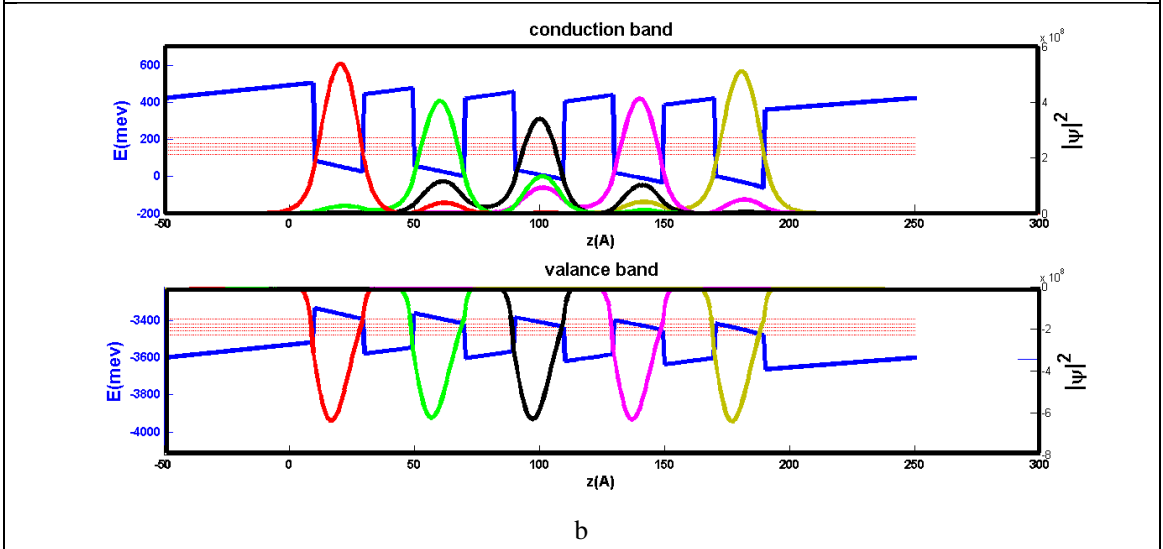

b

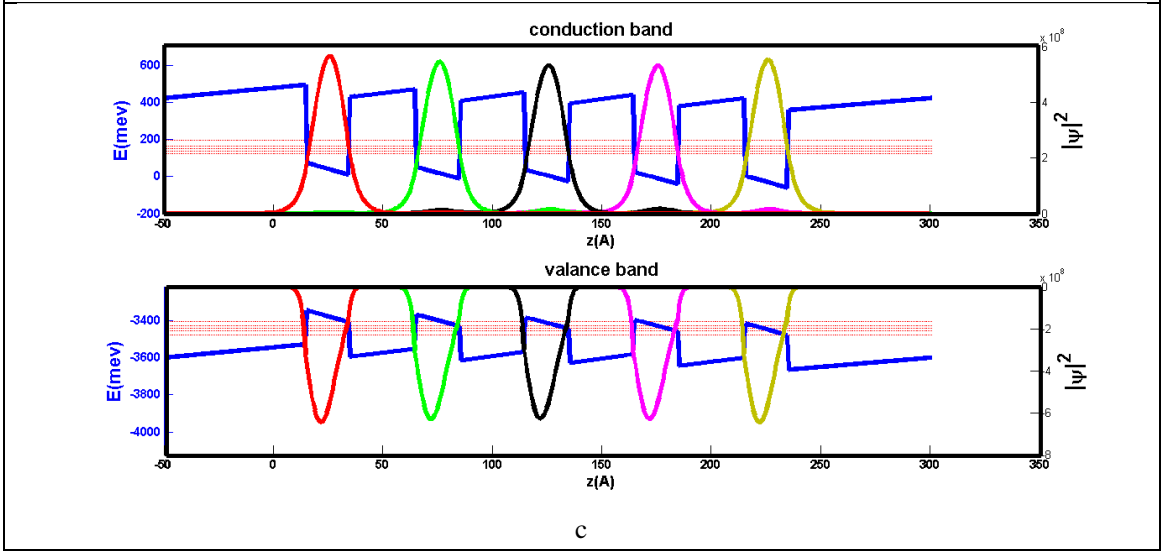

c

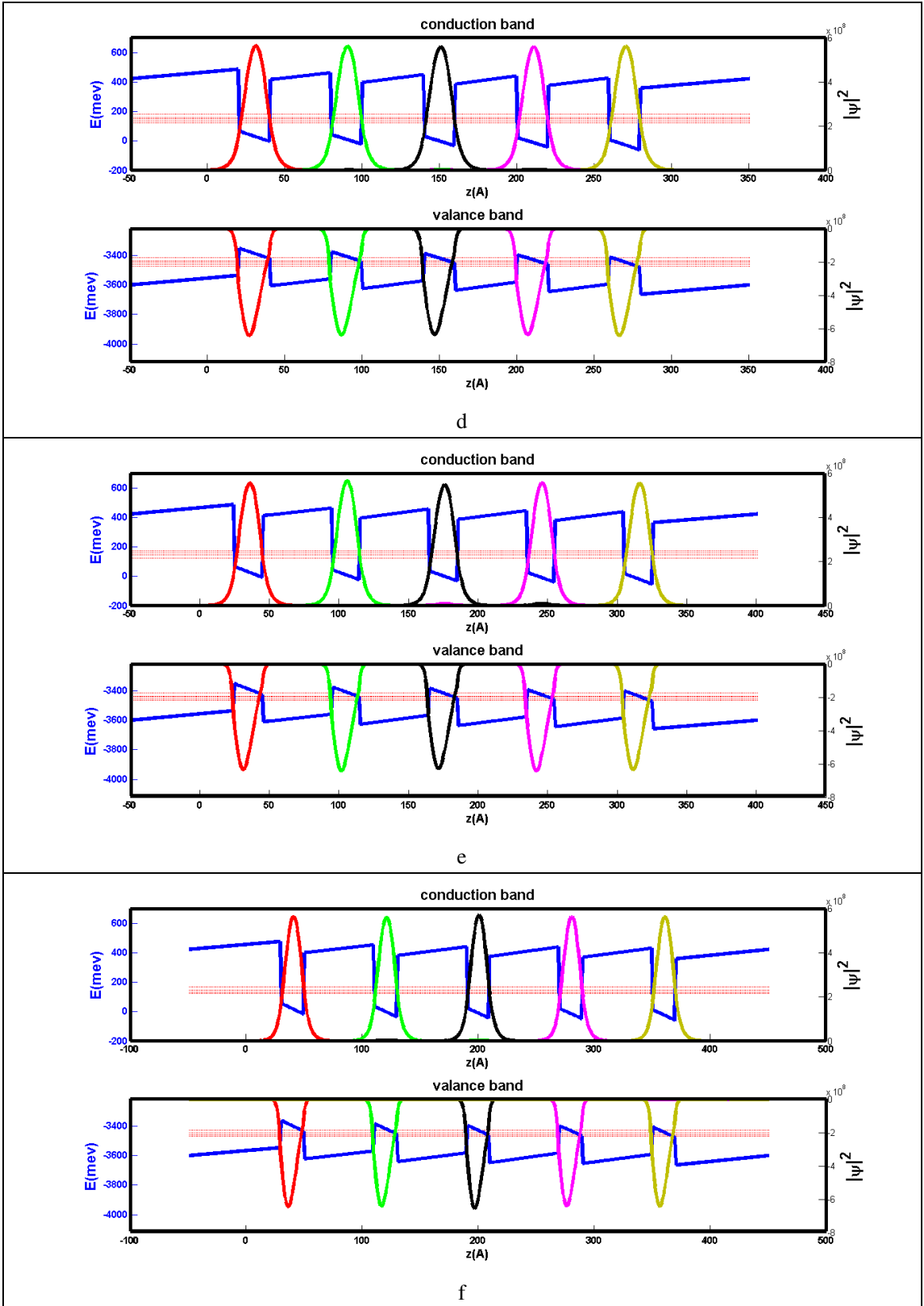

*Fig.10: wave function and band profile for nanowire with different distance of QDiscs of $2L_b = a)10, b)20, c)30, d)40, e)50$ and $f)60\ \dot{A}$, $L_w = 5$. $\rho_D = 1.99 \times 10^{18}\ 1/cm^3$*

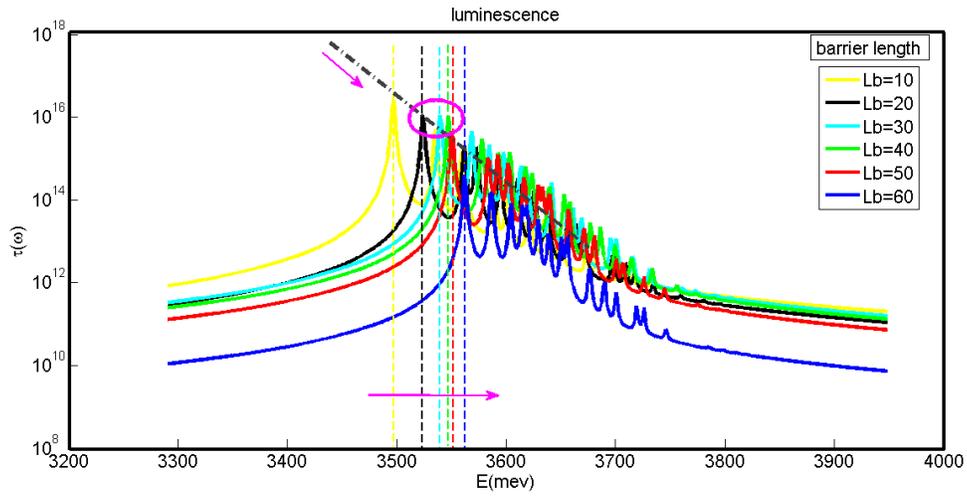

*Fig.11. Luminescence spectrum for 5 GaN QDiscs.* $2L_b = 10,\ 20,\ 30, 40, 50\ and\ 60\ Å$, $L_w = 5Å$, *complete ionized donor as* $\rho_D = 1.99 \times 10^{18}\ 1/cm^3$.

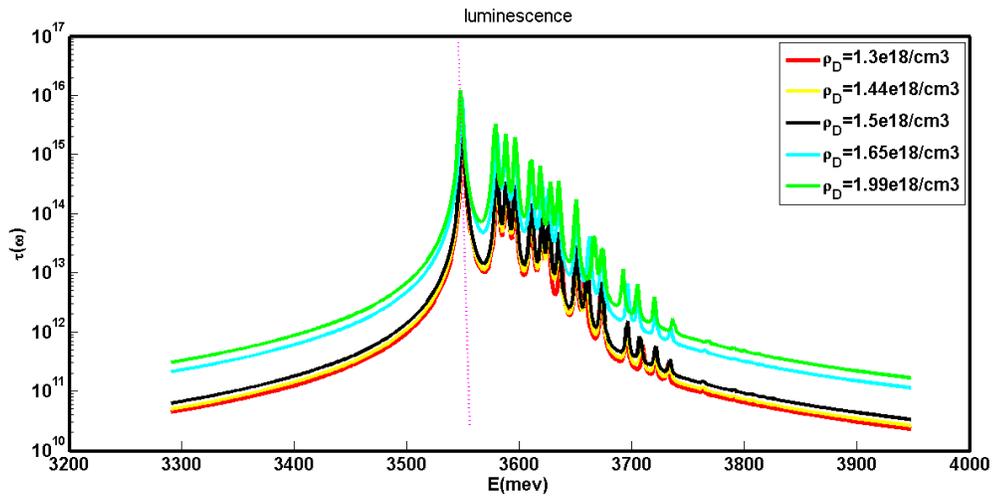

*Fig.12. Luminescence spectrum for 5 GaN QDiscs with different ionized donor level* $L_b = 25Å$, $L_w = 20Å$, , $\rho_D = 1.3 \times 10^{18}, 1.44 \times 10^{18}, 1.5 \times 10^{18}, 1.65 \times 10^{18} and\ 1.99 \times 10^{18}\ 1/cm^3$